\documentclass{aa} 

\usepackage{graphicx}
\usepackage[varg]{txfonts}

\usepackage{hyperref}
\usepackage{natbib}
\bibpunct{(}{)}{;}{a}{}{,}

\def \aap {Astron. Astrophys.}

\def \apjl {Astrophys. J.}

\newcommand{\eq}[1]{\begin{equation} #1 \end{equation}}
\newcommand{\bigz}[1]{\left( #1 \right)}
\newcommand{\bbigz}[1]{\left[ #1 \right]}

\newcommand{\ve} [1]{ {\bf #1}}

\newcommand{\mj}{$^{-1}$} 
\newcommand{\mt}{$^{-3}$} 

\begin{document}

%
%

   \title{Ejection velocities, age, and formation process\\ of SPE meteoroid cluster.}

   \author{
                {David \v{C}apek}
                \and {Pavel Koten}
                \and {Pavel Spurn\'{y}}
                \and {Luk\'{a}\v{s} Shrben\'{y}}
                }

   \institute{Astronomical Institute of the Czech Academy of Sciences,
              Fri\v{c}ova 298,
              251 65 Ond\v{r}ejov,
              Czech Republic\\
              \email{capek@asu.cas.cz}
             }

   \date{Received ??? / Accepted ???}  

\abstract
{
Meteoroid clusters represent a unique opportunity to study processes of meteoroid disruptions in interplanetary space. 
We follow our previous work about the September epsilon Perseid (SPE) meteoroid cluster from 2016 with a detailed analysis of the observed data and cluster formation conditions. 
}
{
Our goal is to determine ejection velocities of the cluster members and SPE's age, as well as to estimate the most probable formation process.
}
{We precisely determined mutual positions and masses of all meteoroids including the errors.
We assumed that the mass-dominated meteoroid is the parent body of the cluster and that the observed positions of meteoroids 
are controlled by the ejection velocities and the action of solar radiation pressure. 
A formula for the dependence of meteoroid ejection velocities on the mutual positions, masses, and cluster age was derived. 
It was assumed that the time at which the initial kinetic energy of all meteoroids reached a minimum value corresponds to the age of the cluster. 
Knowing values and directions of ejection velocities together with meteoroid masses then allowed us to determine the most likely process of cluster formation.
}
{
The meteoroids occupy a volume of $66\times67\times50$~km and are shifted in the antisolar direction by $27$~km relative to the parent meteoroid. 
The age of the cluster is $2.28\pm0.44$~days. The ejection velocities range from $0.13\pm0.05$~m\,s\mj\ to $0.77\pm0.34$~m\,s\mj\ with a mean value of $0.35$~m\,s\mj. 
The ejection velocity directions are inside the cone with an apex angle of $101\pm5^\circ$. The axis of the cone is $\sim45^\circ$ away from the solar direction and $\sim 34^\circ$
away from the mean direction of the flux of small meteoroids' incident on the parent meteoroid.
Formation due to the separation of part of the surface due to very fast rotation is the least likely thing to occur.
We estimate the rotation frequency to be about $2$~Hz and the corresponding stress is several orders of magnitude lower than the predicted strength limit.
It is also difficult to explain the formation of the cluster by an impact of a small meteoroid on the parent body. However, this possibility, although not very likely, cannot be completely ruled out.
The most probable process is the exfoliation due to thermal stresses. Their estimated magnitude is sufficient and the derived ejection velocities are consistent with this process of formation.
}
{}


   \titlerunning{Ejection velocities, age, and formation process of SPE meteoroid cluster.}
   \authorrunning{D. \v{C}apek et al.}

\maketitle

\section{Introduction}
Observations of short-duration meteor outbursts, which are caused by a compact group (``cluster'' in the following text) of meteoroids, are very rare.
Even rarer are the cases of multistation observations, which allow for the precise determination of the relative positions of the individual members, their velocities, and trajectories.  We first briefly review the cases published so far.

Probably the first meteoroid cluster recorded by scientific instruments was described by \citet{HapgoodRothwell1981}. The outburst was observed in 1977 during Perseid shower activity from two stations in England. It consisted of three meteors that were recorded at a time interval of 1.2 s. The authors hypothesized that it was formed by the breakup of an $\sim0.3$~g parent meteoroid by a dust particle. They also estimated the distance from Earth to the breakup point. The lower limit ($\sim 1\,700$~km above Earth's surface) was determined from measurement uncertainties. The upper limit ($0.004$--$0.04$~AU) was determined from assumptions about ejection velocities based on the rotation frequency of such a meteoroid. Impactor parameters and collision probabilities were also estimated. 
However, the question is whether these meteoroids are actually physically related.
Similarly, \citet{Kotenetal2021} observed three Geminid meteors within $0.9$~s and they cannot rule out the possibility that the triplet is only a result of chance.

\citet{PiersHawkes1993} observed five meteors in a fraction of a second over Canada in 1985. Unfortunately, the observation was made from only a single station, so it was not possible to determine the trajectory and velocity of the meteoroids.

The source of rich meteor outbursts is the Leonid shower. \citet{Kinoshitaetal1999} described an observation of more than 100 meteors within $2$~s from Hawaii in 1997. The counts of meteors brighter than the magnitude $+1$, $+2$, $+3$, and $+4$ were $10$, $20$, $40$, and $60$, respectively. Contrary to the single station observation, they determined the atmospheric trajectory of six bright meteors. The cluster dimension was $100$~km in the direction parallel to the velocity vector and $40$~km in the perpendicular direction. Fainter meteors were shifted in the antisolar direction with respect to the brightest ones. They only mentioned the disruption of the parent meteoroid outside Earth's atmosphere as being a cause of the observed outburst, but without a detailed analysis. 

\citet{Watanabeetal2002} reported an observation of 15 meteors in $4$~seconds made in 2001 from Japan. The natural explanation is considered to be the meteoroid disintegration shortly before the encounter with Earth.

\citet{Watanabeetal2003} summarized the abovementioned clusters in the Leonid shower (including one more above Japan in 2002, consisting of 38 meteors within $2$~s) and discussed the origin of the clusters. The spatial extent of the clusters is on the order of $100$~km and they are usually stretched along the velocity vector direction. The authors showed that it is difficult to explain the formation of these clusters by the breakup of the parent meteoroid shortly after ejection from the parent comet, or in the close vicinity of Earth. They suggested that the best explanation is fragmentation in the perihelion, where thermal effects are strongest, and which the meteoroid passed through 6 days before atmospheric entry. Ejection velocities of $20$--$50$~cm\,s\mj\ are required to achieve the appropriate spatial extent during this time. On the other hand, they estimated a spin rate for the parent meteoroid achieved during ejection from the cometary nucleus and considered that the kinetic energy of the rotation transforms into the translation kinetic energy of the fragments. The resulting velocities are on the order of $1$~m\,s\mj, which is close to the abovementioned values. Concerning the mechanism of the fragmentation, they neglected the collisional fragmentation due to a small cross section and preferred disruption due to thermal effects. 


\citet{Kotenetal2017} reported an outburst containing a bright fireball and eight fainter meteors. All of them were observed within $1.5$~s. The event occurred on September~9, 2016 at 23:06:59 UT. A fireball was recorded from five stations of the Czech fireball network. The faint meteors were recorded by the video cameras. Multistation observations allowed for the determination of atmospheric trajectories and heliocentric orbits. The fragments' distances range between 14 and 105 km. All meteoroids belong to the September epsilon Perseid meteor shower (208 SPE). Unlike the other clusters, this one contains a mass-dominated body that apparently corresponds to the parent meteoroid from which the smaller ones were ejected by some process. The smaller meteoroids were shifted in the antisolar direction. It was assumed that this shift was caused by the solar radiation pressure (SRP) and the cluster age was estimated as $2-3$~days.

The present paper follows that of \citet{Kotenetal2017} and focuses on the detailed determination of the cluster age, the ejection velocities of its members, and the determination of the most probable cause of its formation. The text is organized as follows: Section~\ref{secPosMas} deals primarily with quantities that can be inferred directly from the observation -- the masses and relative positions of the fragments. In Section~\ref{ageSec} we use these quantities to derive the cluster age and fragment ejection velocities. Section~\ref{originSec} deals with possible cluster formation processes:  rotational bursting (Section~\ref{rotSec}), impact of small meteoroid (Section~\ref{impactSec}), and thermal stress (Section~\ref{stresSec}). This is followed by a discussion and conclusions. Finally, Appendix~\ref{appPos} describes the determination of mutual positions from the observed data. 
We would like to note that the cluster members are referred to as ``fragments'' in the following text.

\section{Positions and masses of fragments}
\label{secPosMas}
Precise knowledge of the mutual positions of the fragments and their masses, including errors, allows for the determination of the cluster age and initial ejection velocities of individual fragments. 
The cluster was originally described as a fireball, accompanied with eight meteors \citep{Kotenetal2017}. Later we found another faint meteor. The cluster therefore consists of
the main body (fragment 0) and nine smaller fragments (fragments 1 -- 9). The newly discovered fragment has a number of 9.

\begin{figure}[t]
        \centering
    \includegraphics[width=\hsize]{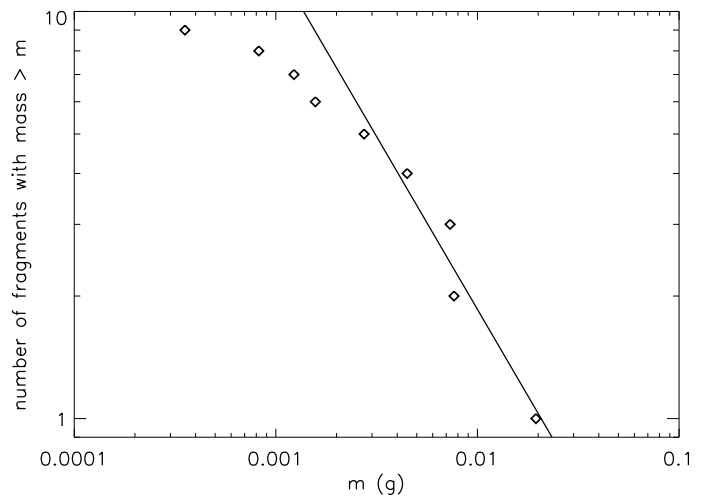}
    \caption{Mass distribution of fragments for the nominal cluster. The line corresponds to $N(>m)\propto m^{-0.85}$, i.e., $D=2.55$, and it is the fit for fragment masses above $0.002$~g. We assume that below this mass, the numbers of observed fragments may not be complete. This is because the corresponding meteors are fainter and may not have been be recorded.
}
    \label{figCumMassDistr}
\end{figure}

%
%
%
\subsection{Masses}
\label{secMasses}
%
%
\renewcommand{\arraystretch}{1.3}
\begin{table*}
\caption{Properties of individual fragments. The photometric preatmospheric mass $m_i$ has two combined standard uncertainty intervals. Size $d_i$ is for density $\rho=1000$~kg\,m\mt. 
The coordinates of corrected observed positions of fragments $\ve{r}^{\rm obs}=(X,Y,Z)$ are expressed with respect to the coordinate system that is connected with the most massive fragment 0. The  $x$-axis points in the antisolar direction, and the $z$-axis points to the north pole of the ecliptics. Lastly, $\sigma^{\rm R}_i$ represents the combined standard uncertainty of positions.}
\label{tabFrag}
\centering
\begin{tabular}{c r@{}l r@{}l rrrr}     
\hline 
 frag. $i$&
 \multicolumn{2}{c}{$m_i$} &
 \multicolumn{2}{c}{$d_i$} &
 \multicolumn{1}{c}{$X_i$} &
 \multicolumn{1}{c}{$Y_i$} &
 \multicolumn{1}{c}{$Z_i$} &
 \multicolumn{1}{c}{$\sigma^{\rm R}_i$}\\
 &
 \multicolumn{2}{c}{($10^{-3}$g)} &
 \multicolumn{2}{c}{(mm)} &
 \multicolumn{1}{c}{(km)} &
 \multicolumn{1}{c}{(km)} &
 \multicolumn{1}{c}{(km)} &
 \multicolumn{1}{c}{(km)} \\
\hline
0& 66&.40\raisebox{0.3ex}{$^{+11.70}_{-10.00}$}~g &  50&.2 &   0.000&    0.000&    0.000&    0.032\\
1& 19&.50\raisebox{0.3ex}{$^{+ 8.80}_{- 6.10}$}   &   3&.3 &  42.516&    8.939&  -37.339&    0.167\\
2&  2&.74\raisebox{0.3ex}{$^{+ 1.85}_{- 1.11}$}   &   1&.7 &  26.977&  -11.916&  -12.268&    0.210\\
3&  7&.65\raisebox{0.3ex}{$^{+ 5.25}_{- 2.98}$}   &   2&.4 &  84.139&  -43.757&  -48.710&    0.952\\
4&  7&.32\raisebox{0.3ex}{$^{+ 3.08}_{- 2.69}$}   &   2&.4 &  70.416&  -29.267&  -33.713&    0.231\\
5&  1&.23\raisebox{0.3ex}{$^{+ 0.77}_{- 0.51}$}   &   1&.3 &  54.924&   14.484&  -37.583&    0.612\\
6&  1&.57\raisebox{0.3ex}{$^{+ 1.02}_{- 0.66}$}   &   1&.4 &  46.286&   -5.252&   -2.559&    0.417\\
7&  4&.48\raisebox{0.3ex}{$^{+ 2.31}_{- 1.69}$}   &   2&.0 &  66.155&   -7.757&  -18.498&    0.259\\
8&  0&.82\raisebox{0.3ex}{$^{+ 0.43}_{- 0.31}$}   &   1&.2 &  92.833&  -14.704&  -30.086&    1.439\\
9&  0&.35\raisebox{0.3ex}{$^{+ 0.31}_{- 0.25}$}   &   0&.9 &  70.728&  -52.815&  -52.372&    2.169\\
\hline
\end{tabular}
\end{table*}
\renewcommand{\arraystretch}{1}

The preatmospheric photometric mass of fragment 0 was determined by \citet{ShrbenySpurny2019} as 66.4~g, using the fireball data from Digital Autonomous Fireball Observatories (DAFO) from the Czech part of the European Fireball Network\footnote{``SPE19'' in their Table~2}. 
In our introductory study of this cluster, \citet{Kotenetal2017} used an inappropriate model for luminous efficiency, which led to an $\sim 10\times$ lower mass of fragment 0.
Preatmospheric photometric masses of smaller fragments were determined from video observations \citep[for details, see][]{Kotenetal2017}. 
The mean values of masses $m_i$ with uncertainty intervals are listed in Table~\ref{tabFrag}. The mass range of smaller fragments is from fractions of milligrams to almost 20 milligrams and their total mass is $0.045\pm0.009$~g. The fragment~0 is approximately $1\,500\times$ more massive than the total mass of the rest of fragments and it can therefore be assumed as the parent meteoroid of the cluster. 

The cluster appeared almost in the middle of the field of view, and there was still enough space on both sides without meteors. We therefore believe that we have detected the whole cluster and that it was not cut off by the edge of the field of view. 
This is true for brighter meteors. Meteoroids with masses below $\sim 2\times10^{-3}$~g could only be recorded within $\sim100$~km of the parent meteoroid in the antisolar direction, which is the edge of the field of view of the more sensitive camera.
It is possible that the observed fragments are only part of the ejected material located in the observed area. As the mass of the fragments decreases, both the brightness of the corresponding meteor and the probability that it will be detected by video cameras decreases. It is therefore useful to estimate the mass of this unobserved matter. 
The size (or mass) distribution of the fragments (especially those of impact origin) can be described by different power laws \citep[e.g.,][]{FujiwaraTakagi1988}. The cumulative number of fragments larger than a given size $s$ can be expressed by a power law with the exponent $D$, that is $N(>s)\propto s^{-D}$ or $N(>m)\propto m^{-D/3}$ for fragments with higher than a given mass. It can be easily derived that fragments with masses in the interval $\bigz{m_{\rm min},\,m_{\rm max}}$ have a total mass of

\eq{
\label{mTot}
m_{\rm min, max}^{\rm tot} = C\frac{3}{3-D}\bigz{m_{\rm max}^{\frac{3-D}{3}}-m_{\rm min}^{\frac{3-D}{3}}},
}
where $C$ is a constant. It allows one to determine the total mass of fragments with masses smaller than $m_{\rm min}$:


\eq{
\label{mTotTilde}
m_{\rm <min}^{\rm tot} = m_{\rm min, max}^{\rm tot}\ m_{\rm min}^\frac{3-D}{3} \bigz{m_{\rm max}^{\frac{3-D}{3}}-m_{\rm min}^{\frac{3-D}{3}}}^{-1}.
}
If the numbers of observed fragments are complete for masses greater than some limit $m_{\rm lim}$, we can simply compute $m_{\rm min, max}^{\rm tot} \equiv m_{\rm S, L}^{\rm tot}$ as the sum of the masses of all observed fragments above this limit. We denoted the lowest mass of these fragments by $m_{\rm S}$ and the highest mass by $m_{\rm L}$. The total mass, including the masses of observed fragments and unobserved matter is then the following:

\eq{
\label{mTotTot}
m^{\rm tot} = m_{\rm S, L}^{\rm tot}\bbigz{1+\ m_{\rm S}^\frac{3-D}{3} \bigz{m_{\rm L}^{\frac{3-D}{3}}-m_{\rm S}^{\frac{3-D}{3}}}^{-1}}.
}
We assume that the numbers of fragments are complete above $m_{\rm lim}=2\times 10^{-3}$~g, and we only interpolated the power function through the values for higher masses. In this case, $N(>m)\propto m^{-0.85}$, which corresponds to $D = 2.55$ (see Figure~\ref{figCumMassDistr}). 
Similar exponents were derived for the mass distribution of fragments from hypervelocity cratering experiments for targets composed of basalt, granite, ice, and sandstone \citep{Gaultetal1963, Horz1969, Fujiwaraetal1977, Cintalaetal1985, Buhletal2014}. However, because it was interpolated using only five points, no strong conclusions can be made from it.
The sum of the masses of fragments that are greater than $2\times 10^{-3}$~g is $m_{\rm S, L}^{\rm tot}=m_1+m_2+m_3+m_4+m_7 = 41.69\times 10^{-3}$\,g  (see Table 1). This value can be used in (\ref{mTotTot}) together with $m_{\rm S}=m_2$ and $m_{\rm L}=m_1$. The total ejected mass\footnote{Unfortunately, it was not possible to determine the uncertainty using the Monte Carlo method (see Section~\ref{clusterClones}), because for about 20\% of the cluster clones, the fit leads to $D\geq 3$.} is then $0.16$~g. The observed fragments represent only 27\% of the total ejected mass.
 The value of $0.16$~g can be used as an upper estimate of the total ejected mass. The actual distribution may have a smaller slope for lower masses and a nonzero cutoff mass.



\subsection{Mutual positions}
Video observations of the cluster enables the luminous trajectory of all fragments and their positions at a specified time to be determined. The positions of fragments  that were actually
observed differ from those above the atmosphere and out of Earth's gravitational field. They were affected by the following: (i) deceleration in the atmosphere before the beginning of ablation and (ii) the influence of the inhomogeneous gravitational field of Earth. The positions were corrected for those effects even though they are less than uncertainty intervals of $\ve{r}^{\rm obs}$. The resulting corrected observed positions $\ve{r}^{\rm obs}$ of fragments can be seen in Table~\ref{tabFrag}. The coordinate system is connected with most massive fragment 0, x-axis points in the antisolar direction, and z-axis  points to the north pole of the ecliptics. Appendix~\ref{appPos} describes the determination of $\ve{r}^{\rm obs}$ in detail. For better readability, we use $\ve{r}^{\rm obs}$ for the observed positions without the term ``corrected'' in the following text.

We can see that the smaller fragments occupy a space of $66 \times 67 \times 50$~km, 
which is shifted relative to the largest fragment by $27$~km in the opposite direction to the Sun. A similar shift of smaller meteoroids relative to larger ones in the antisolar direction was also described by \citet{Kinoshitaetal1999}.
A visualization of the positions of the fragments in space can be seen at \url{https://www.asu.cas.cz/~capek/spe2016pos.mp4}


\subsection{Physical properties}
\label{secPhys}
Let us try to summarize our knowledge of the physical parameters of the SPE material which are important for estimating the age of the cluster and its origin. 
\citet{ShrbenySpurny2019} studied 25 SPE fireballs and found that their material is very similar to that of Perseids in terms of ablation ability and strength. 
They determined the tensile strength as the dynamic pressure for the point of fragmentation. This point corresponds to the flare on the light curve. 
The dynamic pressures were $0.019$--$0.094$~MPa with a mean value of $0.044\pm0.022$~MPa. The meteoroid that is the parent body of the studied cluster corresponds 
to the fireball ``SPE19'' in their Table~2 and it has the highest dynamic pressure, $0.094$~MPa. We adopted this value as the tensile strength.

The density and porosity were not determined for this meteoroid stream.  However, \citet{ShrbenySpurny2019} show that 
the material of SPE has similar properties as the material of Perseids and Orionids. 
\citet{BabadzhanovKokhirova2009} estimated the bulk density of Perseid and Orionid meteoroids to be $1200\pm200$~kg\,m\mt\ and $900\pm500$~kg\,m\mt, respectively. 
For porosities, they obtained values of $45$\% and $62$\%. 
\citet{Narziev2019} determined the density of Perseid and Orionid material to be $1000\pm200$~kg\,m\mt\ and $500\pm200$~kg\,m\mt, respectively. 
For porosities, he obtained values of $57$\% and $79$\%. 

For SPE meteoroids, we adopted the results of \citet{Narziev2019} for Perseids. The density is $\rho=1000$~kg\,m\mt, with a possible range of values from $800$--$1200$~kg\,m\mt\ and a porosity of $\sim60$\%.


\subsection{Cluster clones}
\label{clusterClones}
The masses of the fragments and their mutual positions are key quantities for the determination of the cluster age and ejection velocities. They are, however, determined with some uncertainties. We thus consider a ``nominal cluster'' with fragments' masses and positions according to Table~\ref{tabFrag}. Moreover, 10\,000 cluster clones were generated with a random distribution of masses and positions.

Let $j$ denote the index of the cluster clone and $j=0$ represent the nominal cluster. Then the observed position of the $i-$th fragment of the $j-$th cluster clone is as follows:
\eq{
\ve{r}^j_i=\ve{r}^{\rm obs}_i+\ve{e}_{\rm rnd}\,\delta r^j_i,
}
where $\ve{e}_{\rm rnd}$ is a unit vector in the random direction. Random numbers $\delta r^j_i$ have a Gauss distribution with a standard deviation equal to $\sigma^{\rm R}_i$, according to Table~\ref{tabFrag}.

We consider that masses of fragments have a log-normal distribution. The standard deviation $\sigma^{\rm m}_i$ of masses of the $i-$th fragment can be estimated as

\eq{
\sigma^{\rm m}_i = \frac{1}{4}\ln\bigz{\frac{m^+_i}{m_i^-}},
}
where $m_i^+$ and $m_i^-$ are the upper and lower bound of the mass uncertainty interval (Table~\ref{tabFrag}). The mean value of the distribution is equal to $m_i$.

%
%
%
%

\section{Cluster age and ejection velocities}
\label{ageSec}
\citet{Kotenetal2017} show that all fragments are shifted in the antisolar direction with respect to the most massive fragment 0.
If we assume that the cluster was formed by mass separation from the parent meteoroid, then such an arrangement indicates the action of SRP. 

Let us assume that the parent meteoroid disintegrates into $N$ fragments at the position $\ve{r}^{\rm dis}$ in time $t=0$. Fragments have masses $m_i$ and initial (ejection) 
velocities $\ve{v}_i^{\rm ej}$. The motion of the fragments is affected by SRP. On timescales much shorter than the orbital period, the position $\ve{r}_i$ of each fragment can be described as

\eq{ \ve{r}_i(t)=\frac{1}{2}\ve{a}_it^2+\ve{v}_i^{\rm ej}t+\ve{r}^{\rm dis} \label{rvst}}
for $i=0,\dots,N-1$. Here $\ve{a}_i$ represents an acceleration of the $i-$th fragment due to SRP, which can be expressed as


\eq{\ve{a}_i = -\frac{3AL_{\sun}Q_{\rm pr}}{4\pi c r_{\rm hel}^2}\rho^{-2/3} m_i^{-1/3} \ve{e}_\odot = a_i\,\ve{e}_\odot, \label{aSRP}}
where $L_{\sun}$ is the solar luminosity, $c$ is the speed of light, $r_{\rm hel}$ is the heliocentric distance, $m_i$ is the fragment mass, and $\ve{e}_\odot$ is a unit vector toward the Sun. The radiation pressure efficiency factor $Q_{\rm pr}=1$ and density $\rho$ is assumed to be the same for all fragments. The shape parameter $A_i$ is defined similarly as in \citet{Ceplechaetal1998} as $A=S/V^{2/3}$, where $S$ is an area of the silhouette of the fragment and $V$ is its volume. This parameter depends on the fragments' shape, for example $A\simeq 1.2$ for spherical shapes and $A=1.5$ for cubes. Although the fragment shapes are certainly not the same, we chose $A_i=1.6$.

Let the cluster hit Earth's atmosphere at time $t=T$. The age of the cluster is therefore $T$ and the observation of corresponding meteors reveals the masses $m_i$ of fragments and their positions $\ve{r}_i^{\rm obs} \equiv \ve{r}_i(T)$. Equations (\ref{rvst}) can be rewritten for time $T$ as follows:

\eq{ \ve{r}_i^{\rm obs}-\ve{r}_0^{\rm obs} = \frac{1}{2}\bigz{\ve{a}_i-\ve{a}_0}T^2 + \bigz{\ve{v}_i^{\rm ej}-\ve{v}_0^{\rm ej}}T, \label{rvstT}}
for $i=1,\dots,N-1$. Here, the equation 

\eq{ \ve{r}_0^{\rm obs} = \frac{1}{2}\ve{a}_0T^2 + \ve{v}_0^{\rm ej}T + \ve{r}^{\rm dis}}
was subtracted from the rest of equations. The left-hand sides are known because of the observation and represent the relative positions of fragments $i\neq 0$ with respect to the fragment $i=0$. The right-hand sides contain $3N$ unknown components of ejection velocities and an unknown age for the cluster $T$. Therefore (\ref{rvstT}) is a system of $3N-3$ equations for $3N+1$ unknowns. In general, it does not have  a unique solution and some additional assumptions are necessary.

Fortunately, the cluster has dominant fragment $i=0$ with $m_0\gg m_{i\neq0}$. Table~\ref{tabFrag} shows that the mass of fragment $i=0$ is at least three orders of magnitude higher than the mass of any other fragment. It is therefore reasonable to consider that the ejection velocity of this fragment is negligible compared to ejection velocities of smaller fragments:
\eq{\ve{v}_0^{\rm ej} \ll \ve{v}_{i\neq 0}^{\rm ej}.}
With this simplification, the ejection velocities of smaller fragments ($i=1,\dots N-1$) can be expressed from (\ref{rvstT}) as 

\eq{\ve{v}_i^{\rm ej} \doteq \frac{\ve{r}_i^{\rm obs}-\ve{r}_0^{\rm obs}}{T} - T\frac{\ve{a}_i-\ve{a}_0}{2}. \label{viej}}
The only unknown quantity here is the cluster age $T$.

%
%
\begin{figure*}[t]
    \centering
        \begin{tabular}{l@{\hspace{5mm}}r}
                        \includegraphics[height=5.7cm]{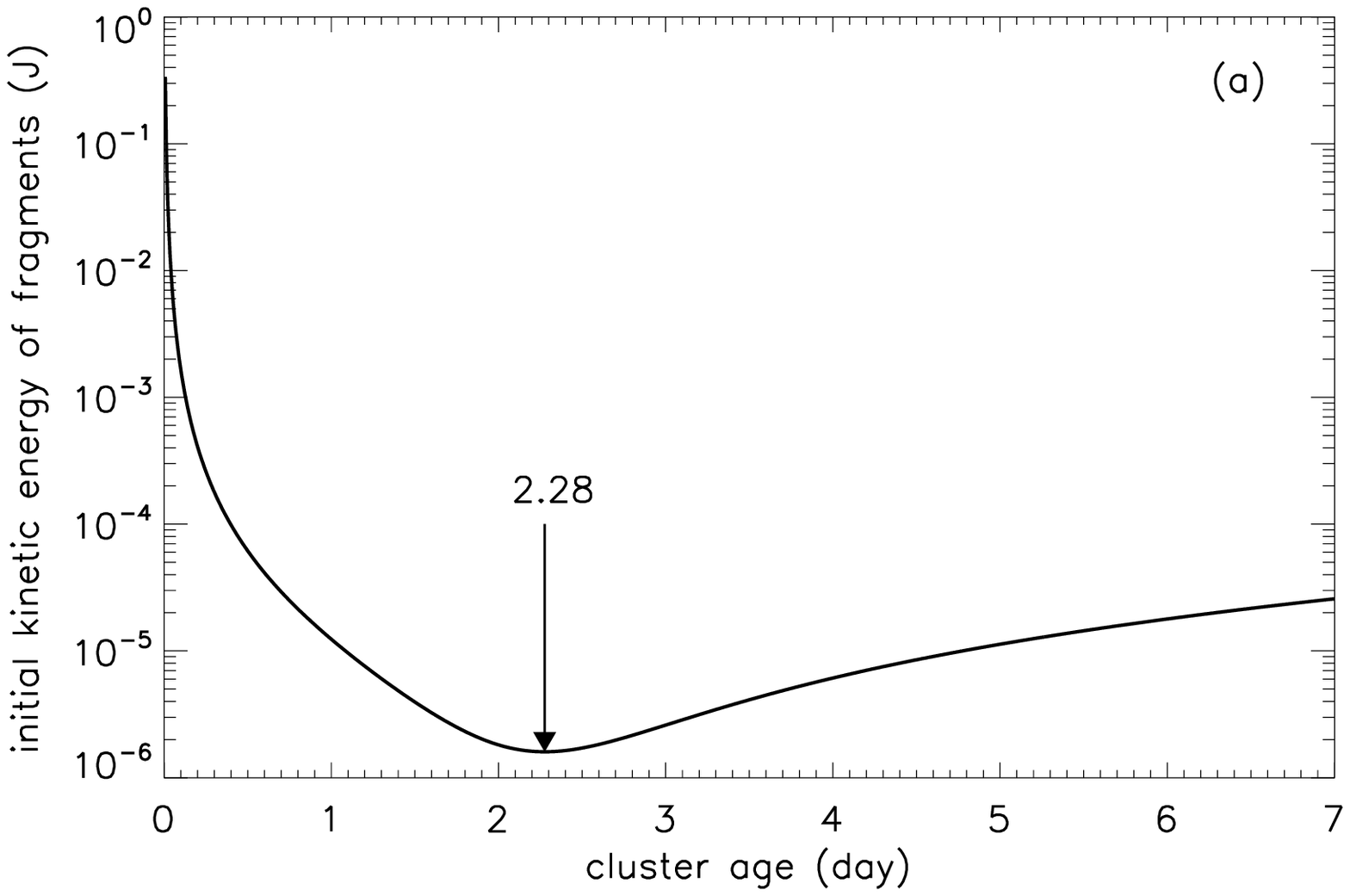}&
                        \includegraphics[height=5.7cm]{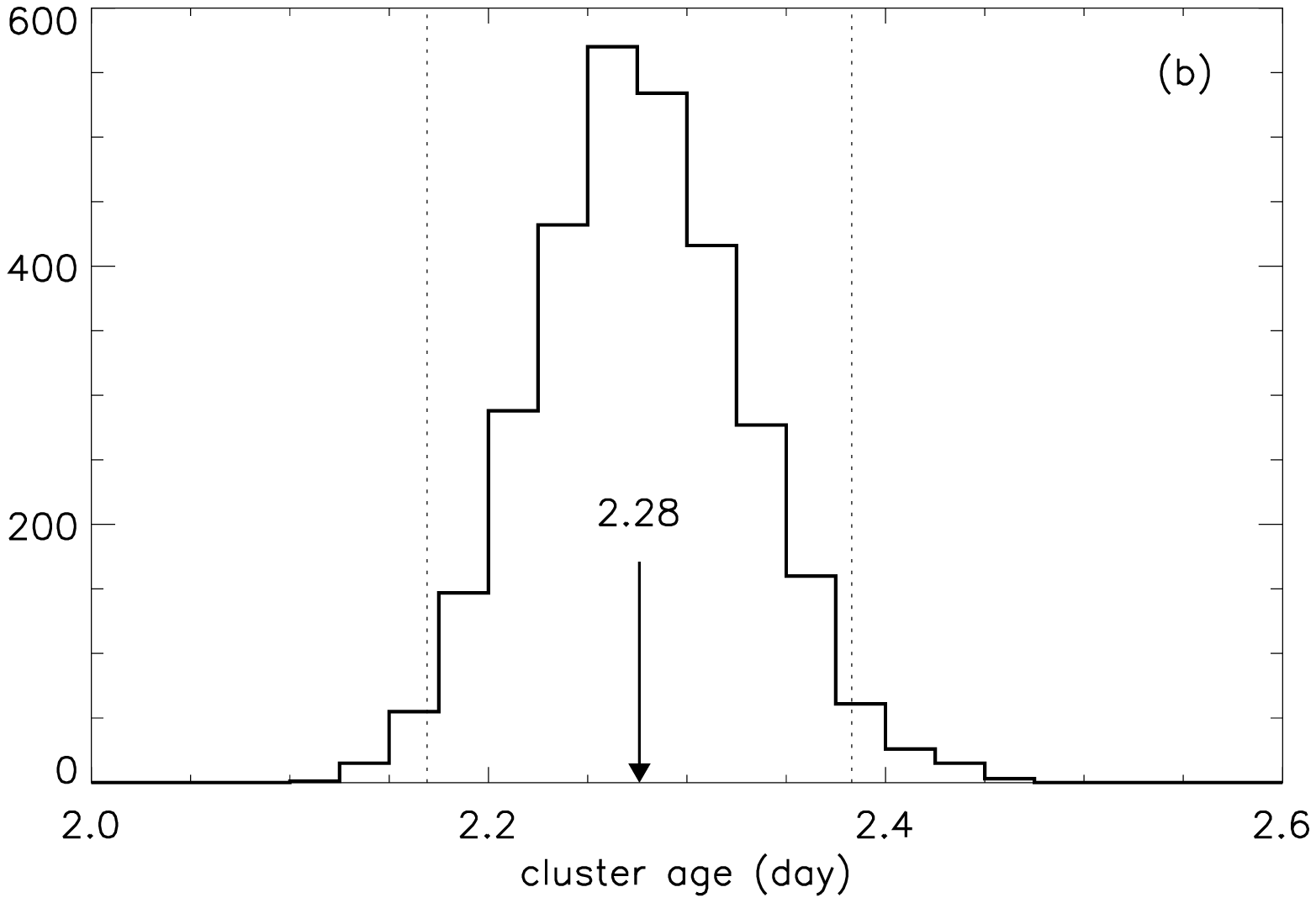}\\
                \end{tabular}
    \caption{Initial kinetic energy and the age of the cluster. (a) Initial kinetic energy of fragments as a function of the age of the nominal cluster. 
                  The position of the minimum, indicated by the arrow, corresponds to the actual age of the cluster, which is 2.28 days.
              (b) Histogram of ages for cluster clones. The arrow corresponds to the mean value and dotted lines represent the $\pm 2\sigma$ interval.}
        \label{figEkVej}
\end{figure*}           

\begin{figure*}[t]
    \centering
        \begin{tabular}{l@{\hspace{5mm}}r}
                        \includegraphics[height=5.7cm]{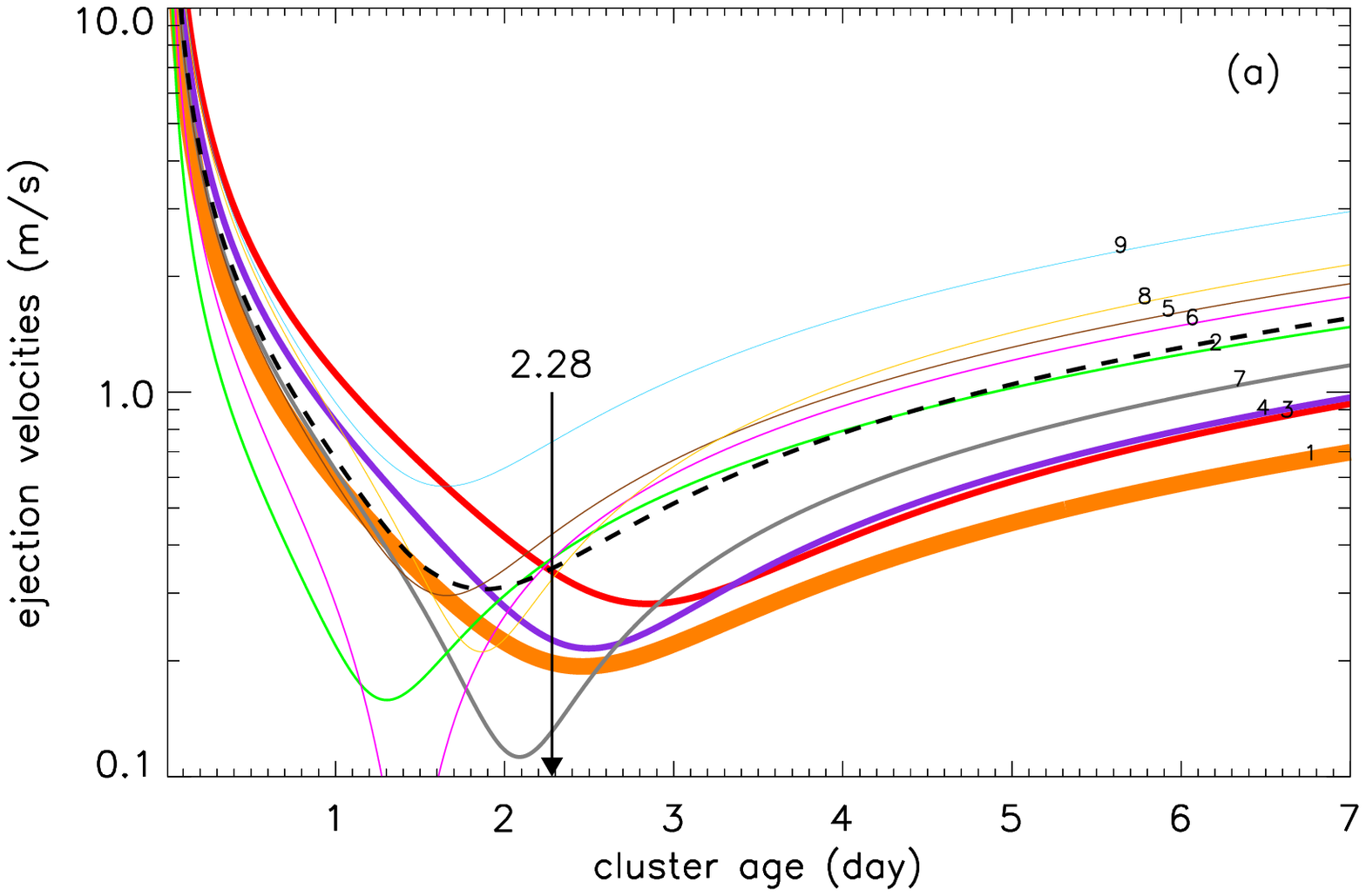}&
                        \includegraphics[height=5.7cm]{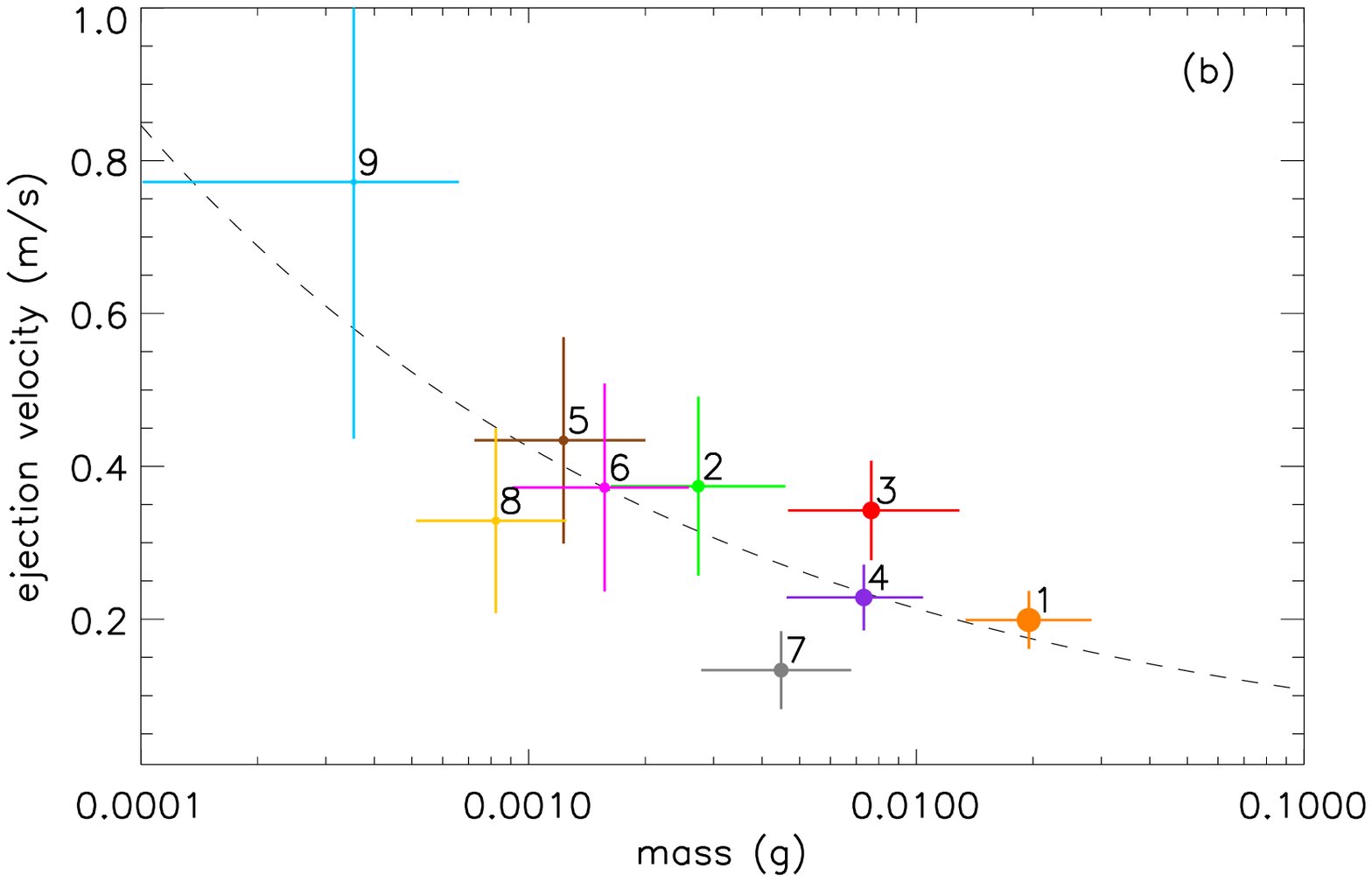}\\
                \end{tabular}
    \caption{Ejection velocity of fragments from the nominal cluster. (a) Ejection velocity as a function of the cluster age.
                  Each curve is labeled with the number of the corresponding fragment and the dashed line represents the mean value.
                         (b) Ejection velocity as a function of mass. The error bars indicate two standard combined uncertainties per side from each point.
                                The dashed curve is the fit by function $v\propto m^{-0.3}$.}
    \label{figmv}
\end{figure*}

\subsection{Cluster age}
The basic assumption for dating the formation of a cluster is the action of SRP. 
A very rough estimation, though the simplest, can be performed under the assumption that the ejection velocities are zero. From (\ref{rvstT}), the age can be estimated for each fragment as 

$$T_i=\sqrt{\frac{2\bigz{\ve{r}^{\rm obs}_i-\ve{r}^{\rm obs}_0}\cdot\ve{e}_\odot}{\bigz{\ve{a}_i-\ve{a}_0}\cdot\ve{e}_\odot}}.$$
The actual ejection velocities are never zero and each fragment reaches the minimum distance from the most massive fragment~0 at different time $T_i$. \citet{Kotenetal2017} determined these times in the interval between (1.4--3) days with a median value of $2.1$~day.

Our current method is based on the assumption that the total kinetic energy of fragments at the cluster formation time $E_{\rm k}^{\rm ej}$ has a minimum value:

\eq{E_{\rm k}^{\rm ej}(T)=\min\limits_t E_{\rm k}^{\rm ej}(t), \label{minEk}}
where

\eq{E_{\rm k}^{\rm ej}=\frac{1}{2}\sum\limits_{i=0}^{N-1}m_i\,\ve{v}_i^{\rm ej} \cdot \ve{v}_i^{\rm ej}. \label{Ekej}}
After substitution of (\ref{viej}) into (\ref{Ekej}) with assumption\footnote{This means that $\partial E_{\rm k}^{\rm ej} / \partial t\,(T)=0$.}  (\ref{minEk}), we were able to derive the following expression for the age of the cluster:

\eq{T=\bbigz{\frac{4\sum\limits_{i=0}^{N-1} m_i\bigz{\ve{r}_i^{\rm obs}-\ve{r}_0^{\rm obs}}^2}
                 {\sum\limits_{i=0}^{N-1} m_i\bigz{\ve{a}_i-\ve{a}_0}^2}}^{1/4}.
}
Using this formula, the age for the nominal cluster is $2.28$~days. Figure~\ref{figEkVej}a shows that this really corresponds to the minimum value of $E_{\rm k}^{\rm ej}$.

Since the masses and positions were determined with some uncertainties, the resulting cluster age also has some uncertainty. To estimate it, we determined the age for cluster clones.
(The creation of cluster clones is described in Section~\ref{clusterClones}.)
The mean value is the same as for the nominal cluster and the $2\sigma$ uncertainty interval is $0.11$~days (Figure~\ref{figEkVej}b). However, the overall uncertainty is larger because the result also depends on physical parameters that are not precisely determined.

The density of $1000$~kg\,m\mt\ is not a directly measured quantity, but only our estimate (see Section \ref{secPhys}). We guess a range of possible fragment material densities in the interval $800$~kg\,m\mt\ -- $1200$~kg\,m\mt. For extreme values of this interval, the age of the cluster is $2.12$~days and $2.42$~days. Another source of uncertainty is the luminous efficiency. We assume that due to this uncertainty, all photometric masses can be up to half or twice the values. Corresponding cluster age extremes are $2.03$~days and $2.56$~days. The last uncertain parameter is the shape factor $A$, which in our model can vary from $1.2$ for spheres to $2$. The latter value is valid for plate-like shapes with body axis ratios of 5:4:1 and 6:3:1 \citep{Sommeretal2013}. Corresponding cluster age extremes are $2.04$~days and $2.64$~days.
Taking all the uncertainties mentioned above into account, the resulting cluster age with two standard combined uncertainties is $2.28\pm0.44$~days.


%
%
\renewcommand{\arraystretch}{1.2}
\begin{table}[t]
\caption{Ejection velocities of fragments. The list contains mean values with two standard combined uncertainties.}
\label{tabVel}
\centering
\begin{tabular}{cr@{}l}     
\hline 
 frag.\#  & \multicolumn{2}{c}{velocity}\\
          & \multicolumn{2}{c}{m\,s\mj }\\
\hline
1 & 0.20 & $\pm$0.04 \\
2 & 0.37 & $\pm$0.12 \\
3 & 0.34 & $\pm$0.07 \\
4 & 0.23 & $\pm$0.04 \\
5 & 0.43 & $\pm$0.14 \\
6 & 0.37 & $\pm$0.14 \\
7 & 0.13 & $\pm$0.05 \\
8 & 0.33 & $\pm$0.12 \\
9 & 0.77 & $\pm$0.34 \\
\hline
\end{tabular}
\end{table}
\renewcommand{\arraystretch}{1}

\subsection{Ejection velocities}
\label{secEjVel}
The ejection velocities of the fragments depend on the cluster age according to (\ref{viej}). This dependence for the nominal cluster can be seen in Figure~\ref{figmv}a. For a higher age, SRP would be acting for a longer time and thus the ejection velocities must be higher and more focused in the sunward direction to shift fragments into their observed positions. In the case of a lower age, the fragments would have less time to be shifted by SRP, so the ejection velocities would be higher and aim toward their observed positions. 
For the age of 2.28~days, the ejection velocities of fragments are lower than $1$~m\,s\mj. 
All fragments have velocities in the range $0.1$--$0.8$~m\,s\mj \ (see Table~\ref{tabVel}). Interestingly, similar velocities ($0.2$--$0.5$~m\,s\mj) were also estimated by \citet{Watanabeetal2003} for the cluster observed during Leonid shower activity. 
The uncertainty intervals were determined similarly as for the cluster age -- from photometric masses and position uncertainties (set by cluster clones), and from assumed extreme values for the material density, the luminous efficiency, and the shape parameter.
In Figure~\ref{figmv}b, we can see a clear dependence of the ejection velocity on the fragment mass. This dependence can be approximated as $v\propto m^{-0.3}$. 

\begin{figure*}[t]
        \centering
    \includegraphics[width=10cm]{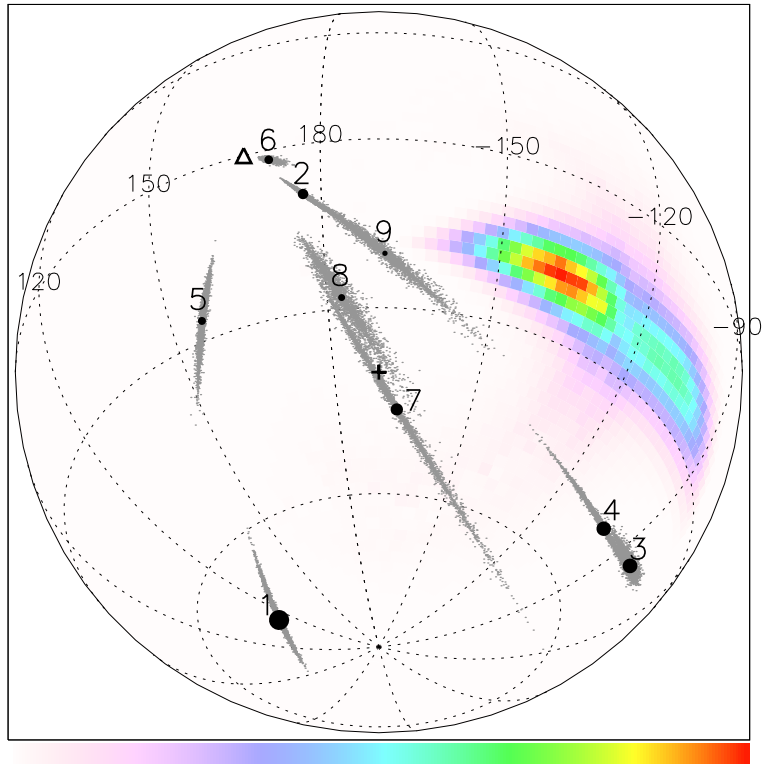}
    \caption{Directions of ejection velocities in the ecliptical coordinate system. The origin of the coordinate system coincides with fragment~0, which is also the parent body of the cluster.
 The black dots with numbers represent the ejection velocity directions of the fragments belonging to the nominal cluster (the size of the dot corresponds to the fragment size) and the black cross in the center of the plot is the mean ejection direction. The gray color is used to show the ejection velocity directions of the cluster clones. For a better view, only 1000 clones are displayed. The triangle shows the position of the Sun. The flux of the small meteoroids' incident on the parent meteoroid is shown by the color scale. This flux was determined using NASA's Meteoroid Engineering Model~3 \citep{Moorheadetal2020}.
}
    \label{figVej}
\end{figure*}

The ejection velocity directions in space are shown in Figure~\ref{figVej}. The black dots represent the ejection velocity directions of the nominal cluster. The small gray points represent the directions for cluster clones. The elongated areas into which they merge then correspond to the uncertainties as to the determination of each direction. The stretching of these uncertainty regions is in the solar -- antisolar directions. This is due to the large uncertainties as to the fragment masses and hence the large scatter in the predicted SRP, which must be balanced by the ejection velocity component in the solar -- antisolar direction. The width of these areas in the direction perpendicular to these directions is given by the uncertainties as to the determination of the relative positions of the fragments. 

\begin{figure}
        \centering
    \includegraphics[width=4cm]{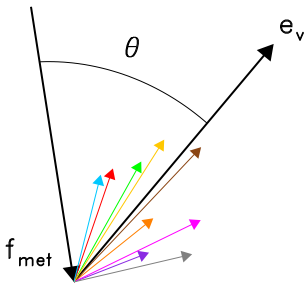}
    \caption{Explanation of the angle $\theta$ between the mean flux of the small meteoroids' $\ve{f_{\rm met}}$ incident on the parent meteoroid and the mean ejection direction $\ve{e_v}$. The colored arrows indicate directions of the ejection velocities of fragments (not to scale).
}
    \label{scatch}
\end{figure}

The directions are relatively close to each other. They are aimed at a limited area, which can be described by a cone with an apex angle of $101\pm 5^\circ$. The mean ejection direction, defined as the axis of this ejection cone, has ecliptic coordinates $l=-171^\circ$ and $b=-40^\circ$, which is $\sim45^\circ$ and far from the solar direction. 
The angle $\theta$ between the mean flux of small meteoroids' incident on the parent meteoroid  \citep[determined using NASA's Meteoroid Engineering Model~3,][]{Moorheadetal2020} and the mean ejection direction is $\sim 34^\circ$ s(ee Figure~\ref{scatch}). 
Due to the high velocity and eccentricity of the SPE orbit, the distribution of radiants of meteoroids hitting the parent meteoroid is different from what is typical on Earth. These radiants are concentrated in a very limited area and it is reasonable to use the mean direction of the flux of the incident meteoroids (see colored area in Figure~\ref{figVej}).    
Interestingly, the directions of the ejection velocities remain the same within assumed intervals of material density $\rho$, luminous efficiency, and shape factor $A$.



\section{Origin of the cluster}
\label{originSec}
Thanks to high quality data, the masses and ejection velocities of all members of the observed cluster as well as its age were determined. This allows us to estimate the most probable cluster formation process.


%
%
%
%

\subsection{Fast rotation}
\label{rotSec}
One of the possible ways how the cluster was formed is that part of the surface was detached due to rapid rotation.
The rotation of the meteoroid in space may be affected by collisions with interplanetary dust particles, as well as by the windmill effect \citep{Paddack1969}.
The windmill effect is caused by the reflection of solar radiation from the irregularly shaped body. It can accelerate the rotation even up to a bursting speed.
The YORP effect \citep{Rubincam2000} is not efficient in this case due to the small size of the parent meteoroid.

Based on a simple model, we can estimate whether this origin is consistent with observed (or derived) properties of the cluster.
Let us assume that the parent meteoroid is a rotating sphere with diameter $d$.
A cube with mass $m_{\rm c}$ is attached to its equator. 
It can be easily derived that the stress at the base of the cube induced by the centrifugal force is as follows: 

\eq{
\sigma = \frac{2 v^2}{d}\rho^{2/3} m_{\rm c}^{1/3},
}
where $v$ is the circumferential velocity. 
We can consider that this velocity is $0.77$~m\,s\mj, that is the maximum ejection velocity. The corresponding spin frequency is $\sim5$~Hz.
The mass of the cube can be assumed as the sum of all fragment masses, which is $0.045$\,g and density $\rho=1000$~kg\,m\mt. 
The resulting stress at the base of the cube is $\sim 85$~Pa.
If we assume an estimated total mass of the ejecta, including those small fragments which were not recorded by the video cameras, 
of $0.16$~g (see Section~\ref{secMasses}), then the stress is $\sim 129$~Pa.
 This stress is $730\times$ lower than the assumed material strength of the meteoroid, 
which was deduced from the fireball behavior in the atmosphere \citep{ShrbenySpurny2019}. 

The rotation speed of the parent meteoroid derived from the ejection velocities is too low to induce stresses on the surface that would be of the same order of magnitude as the strength of the material. Of course, a less idealized situation can be imagined -- an elongated meteoroid shape, a smaller surface on which centrifugal force acts. We can also assume that the circumferential velocity corresponds to the fastest fragment. However, the disproportion between the estimated stress and strength is too large, so this way of cluster formation is very unlikely.

%
%
%
%

\subsection{Collision with small meteoroid}
\label{impactSec}
Another possible  formation process is the impact of a small meteoroid on the parent meteoroid. The impact with less kinetic energy than the threshold kinetic energy \citep{Flynnetal2018, Flynnetal2020} causes a crater instead of a catastrophic disruption of the target. Such a crater is formed in the strength-dominated regime and consists of a bowl-shaped central pit surrounded by a shallower, irregular spallation zone \citep[e.g.,][]{Michikamietal2007,Dufrenseetal2013,Horzetal2020}. The formation of the crater and most of the ejecta occurs in basically two stages. A central depression is formed by excavation and material is ejected at high velocities at an angle of $\sim45^\circ$ with respect to the surface. Larger and slower plate-like spall fragments are then ejected in a direction nearly perpendicular to the surface \citep[e.g.,][]{Hoerthetal2013, Sommeretal2013}.

%
%


In Section \ref{secEjVel} we derive the velocities of the fragments at the moment of their release from the parent meteoroid. These velocities are in the range  $\sim 0.1-0.8$~m\,s\mj. This raises the question of whether the derived velocities and their distribution in space are compatible with the assumption of an impact origin.  The results from hypervelocity impact experiments on solid rock targets produce velocities orders of magnitude larger \citep[e.g.,][]{Gaultetal1963, Flynnetal2018, Flynnetal2020}. Such a simple comparison is not relevant because the ejection velocities depend on the strength of the target, which is much lower than the strength of the targets in the abovementioned experiments.

The expected distribution of the ejecta velocity for the target with properties similar to those of the parent meteoroid can be estimated using the point-source scaling model of \citet{HousenHolsapple2011}. We consider the most appropriate experiment for rescaling to be that of \citet{Housen1992}. He used artificially prepared targets consisting of a mixture of millimeter-sized basalt fragments, iron grit, and fly ash. These targets had a tensile strength of $0.09$~MPa and $0.45$~MPa, and the density was $2600$~kg\,m\mt. The porosity was $23\%$. We used equations (13), (14), and (19) from \citet{HousenHolsapple2011} and substituted the constants corresponding to the experiment of \citet{Housen1992} from their Table~3. For a target with the same tensile strength and density as the parent meteoroid (see Section \ref{secPhys}), it follows that $50\%$ of the ejected mass will have a velocity of less than\footnote{This value is not affected by the velocity and density of the projectile.} $\sim 4$~m\,s\mj. This value is still one order of magnitude higher than the derived ejection velocities of fragments.
Another estimate can be based on the work of \citet{Michikamietal2007}, who performed a series of hypervelocity impact experiments with artificial targets that had different porosities and strengths. According to the curve for the P60 target group (porosity $60\%$, tensile strength $0.056$~MPa, density $1035$~kg\,m\mt) in their Figure~12, we can estimate that $50\%$ of the ejected mass will have a velocity lower than $\sim3$~m\,s\mj. Again, this is a much higher value than for the derived ejection velocities.

A possible explanation for the low ejection velocities of the fragments is that, due to the high porosity of the parent meteoroid, the crater was not formed by excavation but by compaction \citep[e.g.,][]{HousenHolsapple2003}. The same mechanism is used to describe crater formation in material with a similar porosity by \citet{Housenetal1999}. In that case, most of the ejecta had velocities below $1$~m\,s\mj. \citet{Michikamietal2007}, however, argued that this mechanism occurs for materials with even significantly lower strengths.
Another explanation is that the observed fragments, which make up only $\sim 30\%$ of the total ejected mass, are slow spall debris and the rest of the mass consists of small and fast particles. 
Even in this case, however, a sufficiently small velocity $\le0.8$~m\,s\mj\ is not obtained for $30\%$ of the total ejected mass.
It is therefore difficult to explain the discrepancy between very low ejection velocities of the observed fragments and much higher ejection velocities expected in the case of the impact origin of the cluster. 

Another way to make ejection velocities consistent with the impact origin is to allow a different cluster age than that corresponding to the minimum of the kinetic energy of fragments according to condition (\ref{minEk}). Let us assume that a mean ejection velocity of $2$~m\,s\mj\ is consistent with the impact origin of the cluster. This value can be obtained at an age of about $9$ days and higher; ejection velocities would be in the range $0.9-4$~m\,s\mj. However, the directions of the ejection velocities would have to be inside a cone with an apex angle of only $5^\circ$ and the axis of this cone would have to diverge only $2^\circ$ from the Sun to reach the observed positions.  This arrangement is due to the need to balance the longer action of SRP and we consider it highly unlikely.

Reasonable ejection velocities ($0.9-3$~m\,s\mj) can also be achieved at ages of about $10$ hours or less (see Fig.~\ref{figmv}a).
In this case, the action of SRP is of little importance and the ejection velocities are directed toward the observed positions of the individual fragments.
The directions of ejection velocities are inside a cone with an apex angle of about $50^\circ$, and the angle between its axis (which is the mean ejection direction) and the antisolar direction is $\sim 30^\circ$.
The angle $\theta$ between the mean flux of small meteoroids' incident on the parent meteoroid and mean ejection direction is $\sim 90^\circ$ in this case (see Fig.~\ref{scatch}).
This is a slightly unexpected value. Laboratory hypervelocity experiments show that the axis of the ejection cone is usually perpendicular to the surface of the target. In our case, that would mean that it is a very oblique, almost tangential, impact. It could also be an impact from a different and less likely direction than the mean direction of the meteoroid flux. 

The possibility that the cluster was formed by a collision of the parent meteoroid with a small meteoroid about 10 hours (or less) before entering the atmosphere cannot be completely ruled out. Although we do not consider it to be probable.

%
%
%
%
\subsection{Thermal stress}
\label{stresSec}

The surface temperature of the airless bodies of the Solar System is determined by the balance between the absorbed solar radiation, the energy conducted into the deep, and the energy emitted in the form of thermal radiation. The rotation of the body and its morphology periodically changes the flux of incident energy, resulting in temperature variations. If the temperature field in the body is inhomogeneous (and nonlinear),  thermal stresses arise as a result. They can in some situations cause the disintegration of the body, or parts of it. 

This effect was studied for comets or small asteroids or meteoroids falling into the Sun \citep[e.g.,][]{Kuhrt1984, TauberKuhrt1987, ShestakovaTambovtseva1997, TambovtsevaShestakova1999}, small meteoroids orbiting the Sun \citep{CapekVokrouhlicky2010}, or surfaces of asteroids and Moon \citep[e.g.,][]{Delboetal2014, Molaroetal2017}. It is also thought to be the source of the activity of some asteroids \citep{JewittLi2010,Molaroetal2020a}. A breakup due to a ``thermal effect'' has been proposed by \citet{Watanabeetal2003} as the explanation for the meteor outbursts observed during Leonid shower activity.

In order to determine whether the cluster formation by thermal stress is possible, it is first necessary to estimate the magnitude of thermal stresses in the parent meteoroid. A suitable analytical model for this purpose is that of \citet{CapekVokrouhlicky2010}, which allows one to determine the components of the thermal stress tensor in a rotating spherical meteoroid, which is heated by solar radiation. We considered a $5$~cm meteoroid rotating with a frequency of $2$~Hz, and adopted its thermophysical parameters corresponding to carbonaceous chondrites, but with a density of $1000$~kg\,m\mt. 
The spin frequency was very roughly estimated from the assumption that the circumferential velocity corresponds to the mean ejection velocity, which is $0.35$~m\,s\mj. The resulting thermal stress also depends on the angle between the spin axis and the Sun. For a wide range of orientations for the meteoroid's rotation axis (unless it is pointing close to the Sun), the amplitude of the dynamic part of the thermal stress tensor is approximately $0.1$~MPa at the surface\footnote{At the depth equal to the penetration depth of the diurnal temperature wave ($\sim 0.2$~mm), this amplitude drops to about half of this value.}. With the assumed low density and high porosity, a much lower thermal conductivity $K$ can be expected. In that case, the resulting temperature fluctuations and the thermal stress on the surface are even higher\footnote{There are, however, many uncertainties that can affect the result -- for example, the actual thermomechanical parameters of the meteoroid material, the rotational state (free precession instead of rotation about one axis), or the bumpy nonspherical shape.}. The conditions for possible destruction of the material near the surface are therefore met, and processes such as exfoliation of rock flakes along cracks parallel to the surface, as was described for boulders on asteroids, can be expected \citep{Molaroetal2020b}.

\citet{JewittLi2010} explained the observed activity of asteroid (3200)~Phaethon by thermal phenomena near perihelion. They estimated the maximal velocity of particles released during the fracture formation from the transformation of strain energy into kinetic energy. Substituting the material parameters of the parent meteoroid into their equation~(3), it can be estimated that the ejection velocities are no more than a few millimeters per second. We have assumed a thermal expansion of $\alpha=10^{-5}$~K\mj. The temperature variations across the whole surface range from $\sim 50$~K if the rotation axis is pointing toward the Sun to $\sim7$~K for the Sun above the equator. 

\citet{Molaroetal2020b}, using a finite element modeling, explain the episodic ejection of particles from the asteroid (101955)~Bennu by the thermal stress controlled exfoliation of boulders that are on the surface of this body. Their maximum predicted velocity of $\sim 2$~m\,s\mj\ agrees well with the values observed by \citet{Laurettaetal2019}, which range from $\sim 0.05$~m\,s\mj\ to $3.3$~m\,s\mj. Such velocities, however, are produced by boulders with dimensions of a few meters. Figure 7 (left) in \citet{Molaroetal2020b} shows that the maximum velocity of fragments from the surface of a boulder with dimensions of a few tens of centimeters is less than $0.5$~m\,s\mj. 
Thus, the low ejection velocities that we determined for fragments 1 to 9 are consistent with the assumption that the cluster was formed when the surface of the parent meteoroid was disrupted by thermal stress. Moreover, the resulting velocities can be expected to be the sum of the actual ejection velocities (which may be significantly smaller) and the circumferential velocity of the parent meteoroid. 

Thus, the cluster was most likely formed by the destruction of part of the surface of the parent meteoroid due to thermal stresses. Their magnitude is sufficient and the derived ejection velocities of fragments are consistent with this formation process.

\section{Discussion}

As mentioned in Section~\ref{secMasses}, except for the observed bodies, the cluster may contain members that have not been observed. 
We observed no fragments with masses below $0.35\times10^{-3}$~g. This is mainly due to the sensitivity of our instruments. These bodies are strongly affected by SRP, so that a $0.1\times10^{-3}$~g meteoroid with zero ejection velocity would reach $x\simeq270$~km, for example. We assume a cluster age of $2.28$~days and recall that the $x-$axis has its origin in the parent meteoroid and points in the antisolar direction. The observed fragments are located in the interval $x=27-93$~km (Table~\ref{tabFrag}). To be in the same region at the time of observation, the velocity component in the direction toward the Sun would have to be in the range $1.1-1.4$~m\,s\mj. 
Fragments with masses of $(0.35-2)\times10^{-3}$~g were recorded only by the more sensitive camera -- and perhaps not all of them were recorded. The camera's field of view extends to a distance of about $x\simeq 100$~km. At larger distances, the possibility of detecting fragments is minimal and it is therefore possible that the cluster is ``cut'' by the $x\simeq 100$~km boundary for such masses. A fragment with a mass of $0.35\times10^{-3}$~g would certainly escape our attention if the component of its ejection velocity toward the Sun is less than $\sim0.5$~m\,s\mj. We would also not record $2\times10^{-3}$~g fragments with an ejection velocity component in the direction toward the Sun of less than $\sim0.05$~m\,s\mj.
Fragments more massive than $2\times10^{-3}$~g are well detectable by a less sensitive camera whose field of view extends to much greater distances on the $x-$axis. These more massive bodies were probably all detected.

The most likely mechanism of cluster formation seems to be the detachment of a piece of the surface from the parent meteoroid due to thermal stresses. This process leads to very low ejecta velocities, perhaps even lower than the determined ones. We leave a more detailed analysis of this problem for a later study based on \citet{CapekVokrouhlicky2010}, which will be devoted to the effect of thermal stresses on meteoroid populations. In any case, the observed ejection velocities are affected by the circumferential velocity of the rotating parent meteoroid. The degree of this influence depends on from where the material was ejected -- whether that be from the equator or, conversely, from regions around the poles. Of course, we do not know this location. However, the fact that both the largest temperature variations and the largest stress amplitudes are around the equator, rather than at the poles, may provide some clues. The observed ejection velocities can be used to estimate the rotational frequency of the parent meteoroid, not only if we assume formation by rotational fission (Section~\ref{rotSec}), but also if we assume formation by thermal stresses. Thus, we can estimate that the rotational frequency of the parent meteoroid is about $2$~Hz.

To calculate the cluster age, we used assumption (\ref{minEk}) that the initial total kinetic energy of the fragments has a minimum at the time corresponding to this age. However, we have no proof for the validity of this condition. We plan to devote a separate study to its justification in the near future. Let us discuss a situation when this energy is far from its minimum. A very large age implies a precise focus of ejection velocities toward the Sun. A very small age, on the other hand, means focusing the ejection velocities of the fragments toward their observed positions, that is in the antisolar direction. In both cases, there would have to be either a really big coincidence or similar decays would have to occur very frequently. A discussion on this topic can also be found at the end of Section~\ref{impactSec}.

The likelihood of detecting a meteoroid cluster depends on its formation process. From this point of view, processes that lead to large fragment ejection velocities are disfavored.  This is because the fragments scatter outside the observable region much earlier. If one supposes that the ejection velocities of the fragments relative to the parent meteoroid are $10$~m\,s\mj,
in that case, the fragments move away from the parent meteoroid in $3$~hours to about $100$~km. This roughly corresponds to the size of the area covered by the two-station video observation. Of course, this very much depends on the particular geometry, orientation to the relative velocity with respect to Earth, and direction to the Sun. Roughly, we can expect that clusters older than $3$~hours that formed in this way are not detectable by two-station video observations. 
On the other hand, if a cluster is formed in such a way that the ejection velocities are nearly zero, then the maximum age of the observable cluster is given by the effect of SRP on fragments with different masses or shapes. We can then observe clusters that are many days old. However, older clusters whose ejecta are directed toward the Sun would be favored. 


\section{Conclusions}

The most important findings can be summarized in the following points:
\begin{itemize}
\item The cluster consists of a meteoroid with a mass of $66$~g, which is considered as the parent body, and nine other meteoroids with masses in the range $(0.35-19.5)\times10^{-3}$~g. The smaller meteoroids occupy a space of $66\times67\times50$~km and they are shifted by about $27$~km in the antisolar direction with respect to the mass-dominated meteoroid.

\item The resulting (observed) meteoroid positions are controlled by the ejection velocities and subsequent action of solar radiation pressure. 

\item The age of the cluster is $2.28\pm0.44$~days. This age corresponds to the minimum of the cluster formation energy (i.e., initial total kinetic energy of ejected meteoroids). The ejection velocities are in the range $0.13-0.77$~m\,s\mj\ with a mean value of $0.35$~m\,s\mj. The directions of the ejection velocities are inside a cone with an apex angle of $101^\circ$ and the axis of this cone is $45^\circ$ away from the solar direction 
and $\sim 34^\circ$ away from the mean direction of the flux of small meteoroids' incident on the parent meteoroid.
The meteoroid ejection velocities increase with decreasing mass as $v_{\rm ej}\propto m^{-0.3}$. 

\item The formation of the cluster by the separation of a part of the surface due to centrifugal forces is very unlikely.  The rotation frequency estimated from the ejection velocities is about $2$~Hz. The corresponding mechanical stress is several orders of magnitude lower than the strength of the material.


\item Another possible process of cluster formation is the impact of a small meteoroid on the parent meteoroid. In this case, it is difficult to explain small ejection velocities that result for an age of $2.28$ days. They should be at least an order of magnitude higher. 
However, it can be considered that the shift of smaller meteoroids in the antisolar direction with respect to the parent meteoroid is not related to the action of solar radiation pressure.
Ejection velocities would then be sufficient for an age of 10 hours or less. Although less likely, this possibility cannot be completely ruled out.

\item The most likely formation process is exfoliation of the parent meteoroid surface due to thermal stresses. The thermal stresses are sufficiently large and the magnitude of the ejection velocities are consistent with this formation process. A very rough estimate of the spin frequency is $2$~Hz.

\end{itemize}

\begin{acknowledgements}
We are grateful to Petr Pravec for helpful discussions and suggestions.
We thank the anonymous reviewer whose comments helped to improve the text both in terms of content and clarity.
This work was supported by the Grant Agency of the Czech Republic grant 20-10907S and by institutional project RVO:67985815.
\end{acknowledgements}

\bibliographystyle{aa}
\bibliography{speBib}

\begin{appendix}

\section{Mutual positions of fragments}
\label{appPos}
From the video record reduction, the luminous atmospheric trajectory of a fragment is represented by a set of points. Each point is described by time $t$, right ascension $\alpha$, declination $\delta$, and height $h$ (or by geographic coordinates instead of $\alpha$ and $\delta$). These coordinates were transformed into the Cartesian reference frame with the x-axis pointing in the opposite direction than the Sun and the z-axis being perpendicular to the ecliptic plane. The trajectory of $i-$th fragment is therefore described by $\ve{r}_i^j=(x_i^j, y_i^j, z_i^j)$ and time $\tau_i^j$ for $j=1,\dots,n_i$. The number of points per trajectory $n_i$ is from nine (fragments 8 and 9) to 85 (fragment 0).

The meteors were observable in different time intervals and they were never observed all at once. Moreover, their mutual positions were affected by different deceleration in the atmosphere before ablation started and meteors became visible. Another effect that affects their mutual positions is tidal distorsion in Earth's gravitational field. Our goal is therefore to determine the mutual positions of all fragments at a given time with a corrected effect of atmospheric deceleration and tidal distorsion.

At first it is necessary to determine the mean point on the luminous atmospheric trajectory of each fragment: $\ve{r}^{\rm m}_i$, $\tau^{\rm m}_i$. We assumed that the atmospheric motion of the $i-$th fragment can be approximated by a line

\eq{\ve{r}_i^j = \ve{r}_i^{\rm start} + \ve{v}_i\,\tau_i^j.}
The unknown vectors $\ve{r}_i^{\rm start}$ and $\ve{v}_i$ were determined by the least square method. The mean point on the luminous atmospheric trajectory of the $i-$th fragment is as follows:

\eq{
\ve{r}_i^{\rm m} = \ve{r}_i^{\rm start} + \ve{v}_i\,\tau_i^{\rm m},
}
where 

\eq{
\tau_i^{\rm m} = \frac{1}{n_i}\sum\limits_{j=1}^{n_i} \tau_i^j.
}
Mean points of luminous trajectories correspond to different times $\tau_i^{\rm m}$; it is therefore necessary to determine the positions of fragments in one reference time. The reference time was chosen as

\eq{
\tau^{\rm ref} = \frac{1}{n}\sum\limits_{i=1}^{n} \tau_i^{\rm m}.
}
The observed uncorrected positions of fragments' $\ve{r}^{\rm u}_i$ can be expressed as $\ve{r}^{\rm u}_i=\ve{r}_i^{\rm m} + \ve{v}_i \bigz{\tau^{\rm ref}-\tau_i^{\rm m}}$. The atmospheric trajectory of smaller fragments contain a low number of points. Their velocity vectors $\ve{v}_i$ have large errors 
and they are inappropriate for the extrapolation of the trajectory. The velocity vectors may differ due to different initial ejection velocities, an action of SRP to fragments of various masses, an interaction with the atmosphere and the action of tides of Earth's gravitational field. 
These variations are much smaller than the errors. 
We can consider that all velocity vectors are the same and equal to $\ve{v}_0$, which has the smallest error of 

\eq{\ve{v}_i =\ve{v}_0.}
Thus, the uncorrected positions of all fragments in time $\tau^{\rm ref}$ are as follows:

\eq{
\ve{r}^{\rm u}_i=\ve{r}_i^{\rm m} + \ve{v}_0 \bigz{\tau^{\rm ref}-\tau_i^{\rm m}}.
}
However, the motion of the fragments before their observation was affected by (i) deceleration in the atmosphere before the start of ablation and (ii) by tides of Earth's gravitational field and it is necessary to correct these effects.

The positions of fragments above the atmosphere were determined by the backward integration of equations of motion using single body theory without ablation \citep[e.g.,][]{Ceplechaetal1998}. We assumed spherical shapes, density $\rho=1000$~kg\,m\mt, drag coefficient $\Gamma=0.46$, zenith angle of radiant $z_{\rm r}=41.4^\circ$, and the density of the atmosphere according to the NRLMSISE-00 model \citep{Piconeetal2002}. The influence of the tides of Earth's gravitational field was also estimated. We assumed only tides in the radial direction and integrated the motion backward to the distance of $10^6$~km from Earth. The resulting changes $\delta_{\rm atm}$ in mutual fragment positions due to deceleration in the atmosphere before the start of the ablation are lower than $10$~m and the effect of tides $\delta_{\rm tide}$ does not exceed 600~m. 
All these corrections were applied to observed, uncorrected positions according to 

\eq{
\ve{r}^{\rm obs}_i=\ve{r}^{\rm u}_i + \ve{e}_{\rm v} \bigz{\delta_{\rm atm}+\delta_{\rm tide}},
}
where $\ve{e}_{\rm v}$ is the unit vector in the velocity direction, and $\ve{r}^{\rm obs}_i$ are the corrected, observed positions of fragments listed in Table~\ref{tabFrag}.

\end{appendix}

\end{document}